\documentclass[12pt]{iopart}

\usepackage{iopams}  
\usepackage{graphicx}
\begin{document}

\title[Anisotropic flow at the LHC measured with the ALICE detector.]
{Anisotropic flow at the LHC measured with the ALICE detector.}

\author{Raimond Snellings for the ALICE Collaboration.}
\address{
Nikhef National Institute for Subatomic Physics, Amsterdam, 
and 
Utrecht University,
Princetonplein 5,
3584 CC Utrecht,
The Netherlands.}
\ead{R.J.M.Snellings@uu.nl}

\begin{abstract}
The ALICE detector at the LHC recorded first Pb--Pb collisions at
$\sqrt{s_{_{\rm NN}}} = 2.76$ TeV in  November and December of 2010.
We report on the measurements of anisotropic flow for charged and identified particles. 
From the  comparison with measurements at lower energies and with model predictions 
we find that the system created at these collision energies is described well by hydrodynamical model calculations 
and behaves like an almost perfect fluid.
\end{abstract}



\section{Introduction}
Anisotropic flow is an important observable in ultra-relativistic heavy-ion collisions
as it signals the presence of multiple interactions between the constituents of the created matter.
Anisotropic flow has been observed in nucleus--nucleus collisions from 
low energies up to $\sqrt{s_{_{\rm NN}}} = 2.76$~TeV at the Large Hadron Collider (LHC)~\cite{Voloshin:2008dg,Aamodt:2010pa}. 
The azimuthal anisotropic flow is usually characterized by the Fourier coefficients~\cite{Voloshin:1994mz}:
\begin{equation}
v_n = \langle \cos [n(\phi - \Psi_n)]\rangle,
\label{eq:harmonics}
\end{equation}
where $\phi$ is the azimuthal angle of the particle, $\Psi_n$ is the azimuthal angle of the initial state spatial plane of symmetry, 
and $n$ is the order of the harmonic. The second Fourier coefficient $v_2$ is called elliptic flow~\cite{Ollitrault:1992bk}.
Because the magnitude of the anisotropic flow depends strongly on the friction in the created matter, characterized 
by the shear viscosity over entropy density ratio $\eta/s$, the large elliptic flow observed at RHIC provided compelling evidence 
for strongly interacting matter which, in addition, appears to behave like an almost perfect fluid~\cite{Kovtun:2004de}.
However, a precise determination of $\eta/s$ in the partonic fluid is complicated by uncertainties in 
the initial conditions of the collision, the relative contributions from the hadronic and partonic phase, 
and the unknown temperature dependence of $\eta/s$. 
Because of these uncertainties it was not even clear if the elliptic flow would increase 
or decrease when going from RHIC to LHC energies; a measurement of elliptic flow at the LHC was therefore 
one of the most anticipated results.

\section{Integrated Elliptic Flow}
\begin{figure}[h!]
\includegraphics[width=8cm]{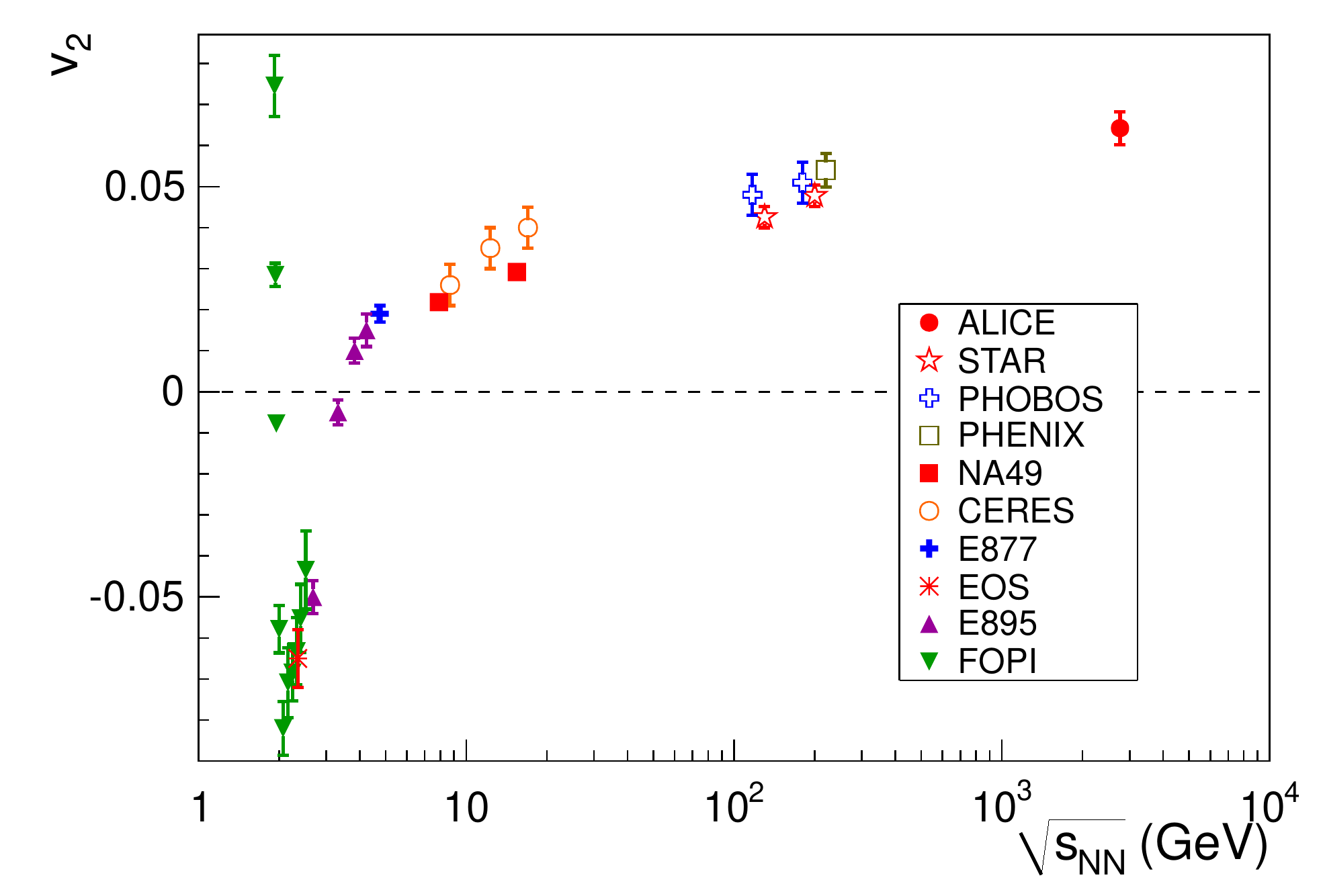}
\includegraphics[width=8cm]{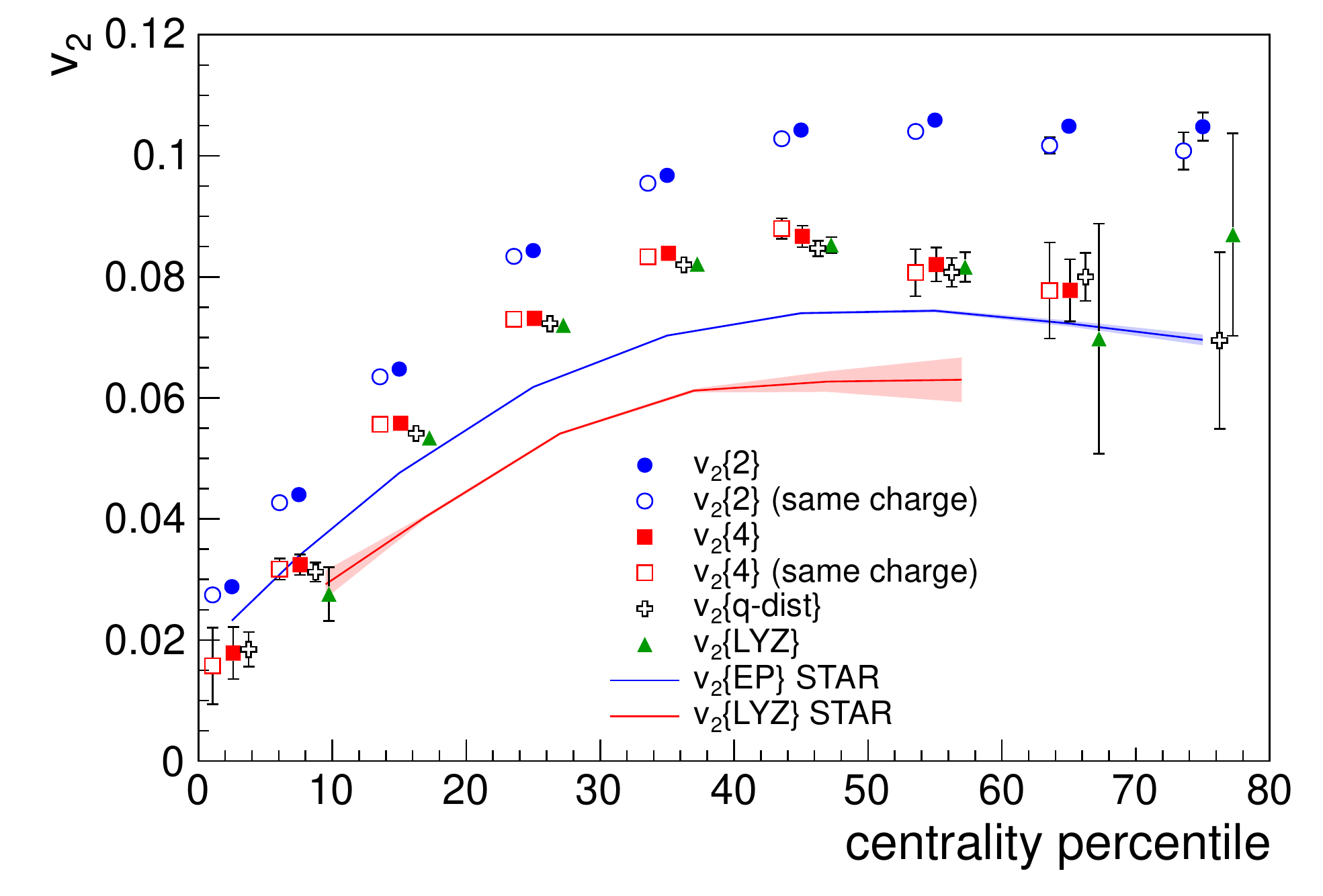}
\caption{
Integrated elliptic flow as a function of collision energy (left) and 
as function of centrality (right). Central (peripheral) collisions correspond to small (large) values of the centrality percentile. 
Figures taken from~\cite{Aamodt:2010pa}.}
\label{fig:figure1} 
\end{figure}

In the left panel of Fig.~\ref{fig:figure1} the ALICE measurement at 2.76 TeV~\cite{Aamodt:2010pa} shows that the integrated elliptic flow of 
charged particles increases by about 30\% compared to flow measured at the highest RHIC energy of 0.2 TeV. 
This result indicates that the hot and dense matter created in these collisions at the LHC still behaves like a fluid with almost 
zero friction, providing strong constraints on the temperature dependence of $\eta/s$.

Experimentally, because the planes of symmetry $\Psi_n$ in Eq.~\ref{eq:harmonics} are not known, the anisotropic flow coefficients are estimated 
from measured correlations between the observed particles. 
The right panel shows the integrated elliptic flow  ($|\eta| < 0.8$ and $0.2 < p_{\rm t} < 5$ GeV/$c$) 
obtained from two- and multi-particle correlations as a function of collision centrality, 
compared to STAR measurements at RHIC. 
Here the elliptic flow estimated from two-particle correlations is denoted by $v_2$\{2\}, while those estimated from multi-particle correlations 
are denoted by $v_2$\{4\}, $v_2$\{{\bf q}-dist\} and $v_2$\{LYZ\}, for the four particle cumulant, fit of the flow {\bf q}-vector distribution 
and the Lee-Yang zeros method, respectively (see~\cite{Aamodt:2010pa} and references therein).

\begin{figure}[htb]
\includegraphics[width=8cm]{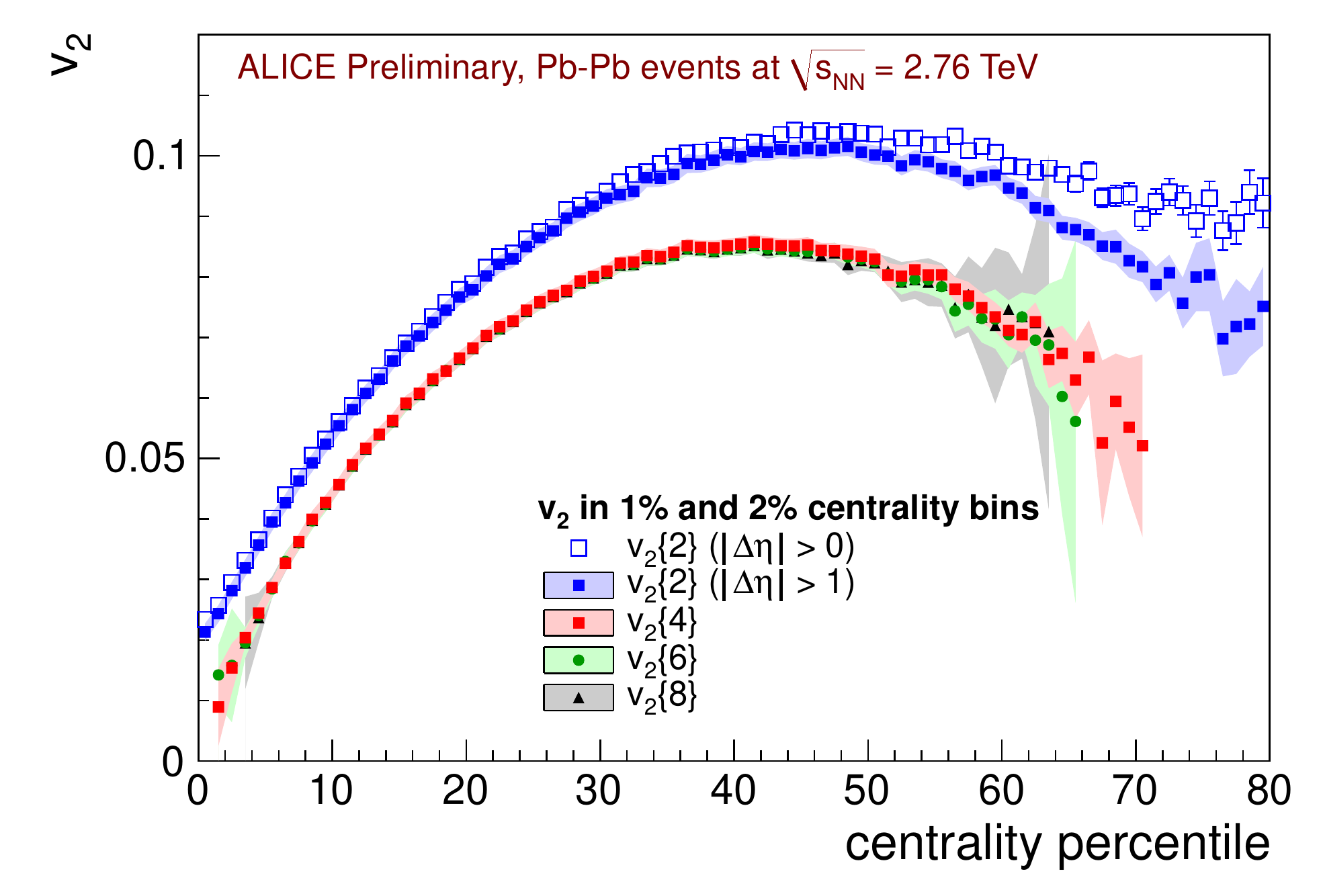}
\includegraphics[width=8cm]{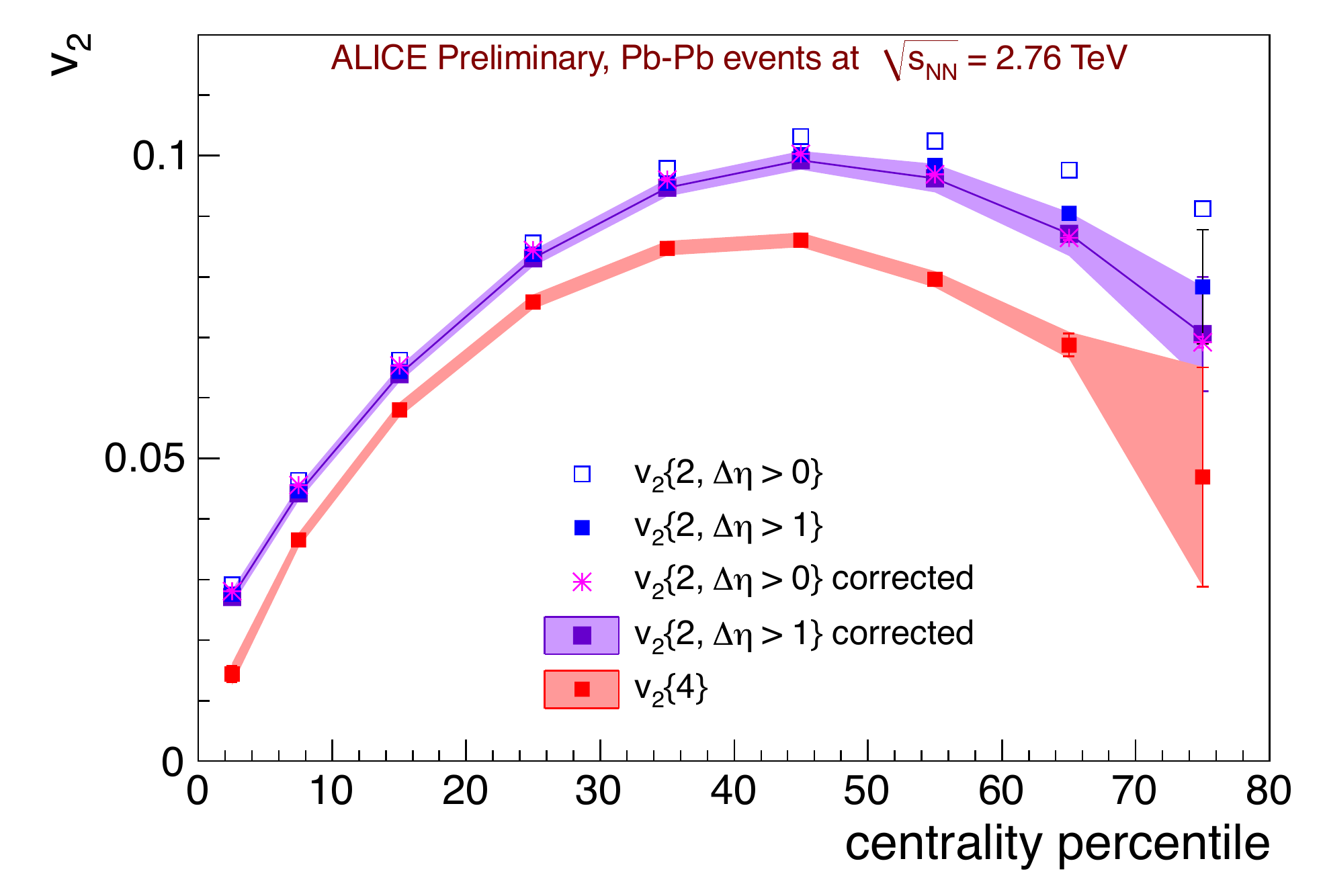}
\caption{
Left: Integrated elliptic flow estimates as a function of collision centrality in 1--2\% bins. 
Figure taken from~\cite{ref:Ante}.
Right: Integrated elliptic flow estimates as a function of collision centrality corrected for nonflow.
The bands show the systematic and statistical uncertainty added in quadrature.
}
\label{fig:figure2} 
\end{figure}
There is a significant difference between flow estimates from two- and multi-particle correlations both at LHC and at RHIC energies.  
This difference is caused by nonflow contributions (these are other sources of azimuthal correlations 
due to jets and resonance decays, for instance) 
and by event-by-event fluctuations in the elliptic flow.  

In the left panel of Fig.~\ref{fig:figure2}, the collision centrality dependence of $v_2$ is plotted in narrow bins (1--2\%), 
to reduce trivial event-by-event fluctuations within a centrality bin. 
In this figure the effect of the nonflow on two-particle estimates is apparent from the difference in $v_2$ 
calculated from particles with $|\Delta\eta| > 0$ and $|\Delta\eta| > 1$.
Also shown are results from 
four-, six-, and eight-particle cumulant estimates which are consistent within uncertainties, 
indicating that the genuine 4-particle, and higher order, nonflow is negligible.  
The flow estimates from the cumulants can therefore be written as~\cite{Voloshin:2008dg}
\begin{equation}
v^2_2\{2\} \approx \bar{v}^2_2 + \sigma_{v_2}^2 + \delta, \;\;\; v^2_2\{4\} \approx \bar{v}^2_2 - \sigma_{v_2}^2, 
 \;\;\; v^2_2\{6\} \approx \bar{v}^2_2 - \sigma_{v_2}^2, 
 \label{eq:cumulants}
\end{equation}
where $\bar{v}_2$ is the event averaged elliptic flow, $\sigma_{v_2}$ the standard deviation of the event-by-event fluctuations 
and $\delta$ the residual nonflow contribution.  Assuming that $\sigma_{v_{2}} \ll \bar{v}_2$, Eq.~\ref{eq:cumulants} is valid up 
to order $\sigma_{v_{2}}^2$.

We use the HIJING event generator (which does not include flow) to estimate $\delta$.   
The right panel of Fig.~\ref{fig:figure2} shows $v_2\{2\}$ for 
$|\Delta\eta| > 0$ and $|\Delta\eta| > 1$ before and after the estimated nonflow correction. 
After this correction both estimates are in good agreement, indicating that HIJING seems to correctly describe the two-particle 
nonflow contribution. 
We currently assign the entire HIJING based nonflow correction for $v_2\{2, |\Delta\eta| > 1\}$ as a conservative estimate of the systematic uncertainty.

\begin{figure}[h!]
\includegraphics[width=8cm]{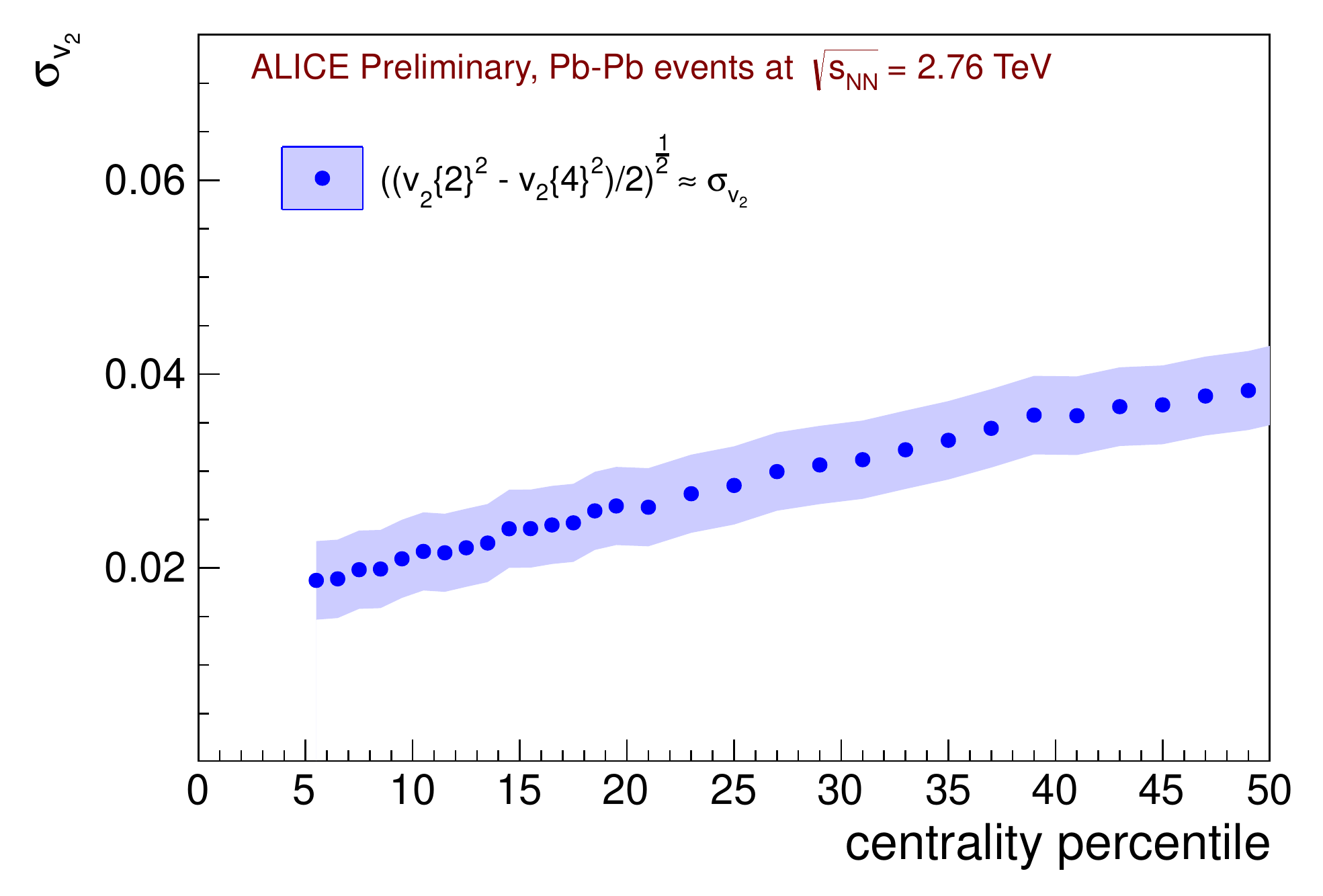}
\includegraphics[width=8cm]{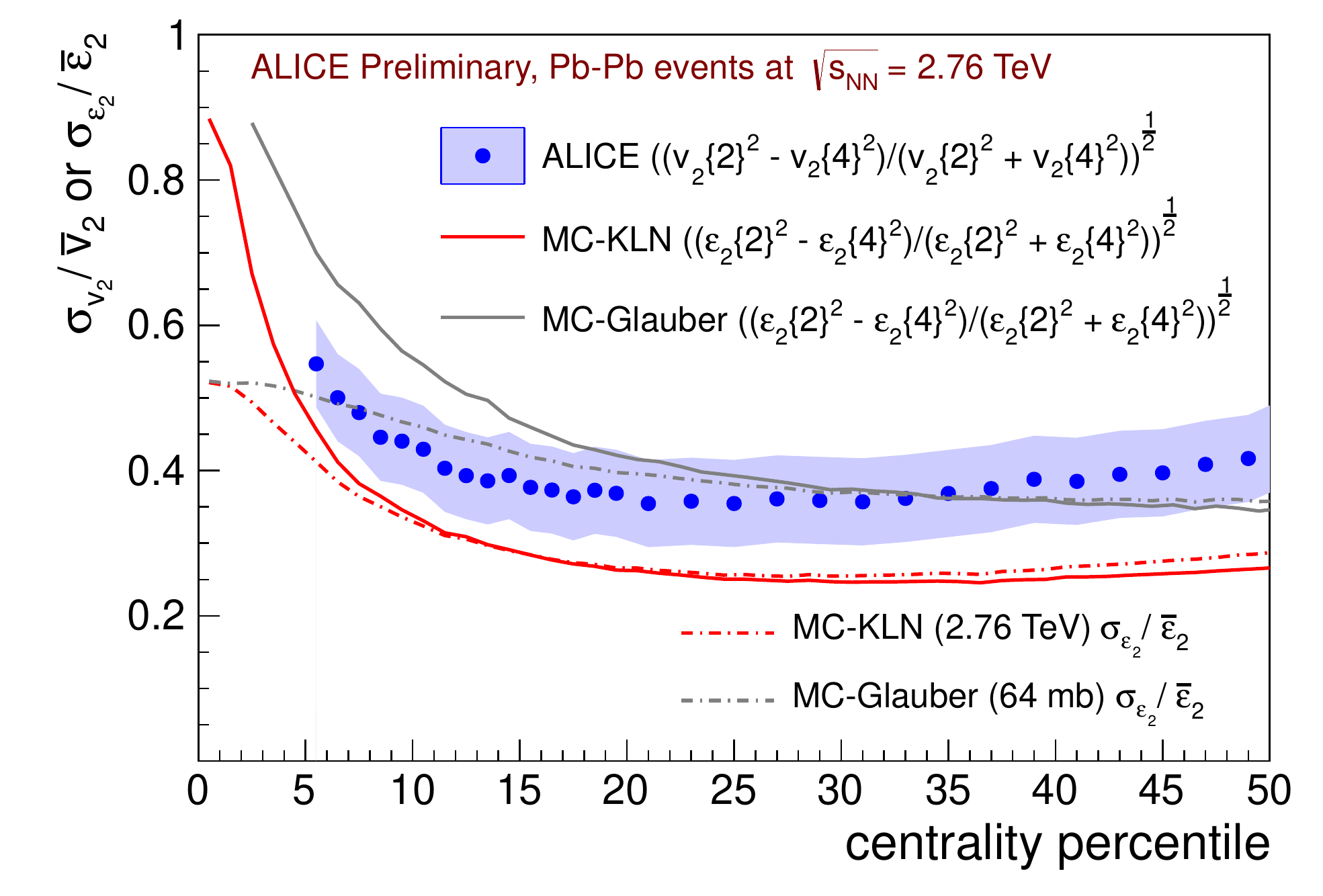}
\caption{
Left: The magnitude of the event-by-event elliptic flow fluctuations versus collision centrality.
Right: Relative event-by-event elliptic flow fluctuations versus collision centrality.}
\label{fig:figure3} 
\end{figure}
Using the nonflow corrected $v_2\{2\}$ and $v_2\{4\}$ we obtain from Eq.~\ref{eq:cumulants} 
\begin{equation}
\bar{v}_2 \approx \sqrt{\frac{v_2^2\{2\} + v_2^2\{4\}}{2}}  \quad \mbox{and} \quad
\sigma_{v_2} \approx \sqrt{\frac{v_2^2\{2\} - v_2^2\{4\}}{2}}.
\label{eq:flucts}
\end{equation} 
The results are plotted in Fig.~\ref{fig:figure3} (left), together with the ratio $\sigma_{v_2}$/$\bar{v}_2$ (right). 
The ratio $\sigma_{v_2}$/$\bar{v}_2$ is found to be large $\sim 40$\%; a similar result has been obtained at RHIC~\cite{Alver:2007qw}.

Because the magnitude of the elliptic flow is proportional to the eccentricity $\varepsilon_2$ of the initial nuclear overlap region 
in the transverse plane, we expect that the event-by-event fluctuation in $v_2$ will be proportional to that in $\varepsilon_2$.
The measured ratio $\sigma_{v_2}/\bar{v}_2$ is compared in Fig.~\ref{fig:figure3} to the ratio $\sigma_{\varepsilon_2}/\bar{\varepsilon}_2$, 
calculated with Eq.~\ref{eq:flucts} from  a MC-KLN~\cite{Drescher:2007ax} and a MC-Glauber model~\cite{Miller:2007ri} (full curves). 
The MC-KLN under-predicts the data whereas the MC-Glauber over-predicts the data for more central collisions.
To investigate to which extent the ratio obtained from Eq.~\ref{eq:flucts}  represents 
$\sigma_{\varepsilon_2}/\bar{\varepsilon}_2$ (and by implication $\sigma_{v_2} / \bar{v}_2$) we have calculated this ratio directly from 
the distributions generated by the two models, that is, without the assumption that $\sigma_{\varepsilon_2} \ll \bar{\varepsilon}_2$.
The result is plotted in the right panel of Fig.~\ref{fig:figure3} for MC-KLN (dotted curve) and MC-Glauber (dot-dashed curve) 
and indicates, in comparison to the full curves, that Eq.~\ref{eq:flucts} breaks down for $\sigma_{\varepsilon_2} /\bar{\varepsilon}_2 \gtrapprox 0.4$. 

\begin{figure}[h!]
\includegraphics[width=8cm]{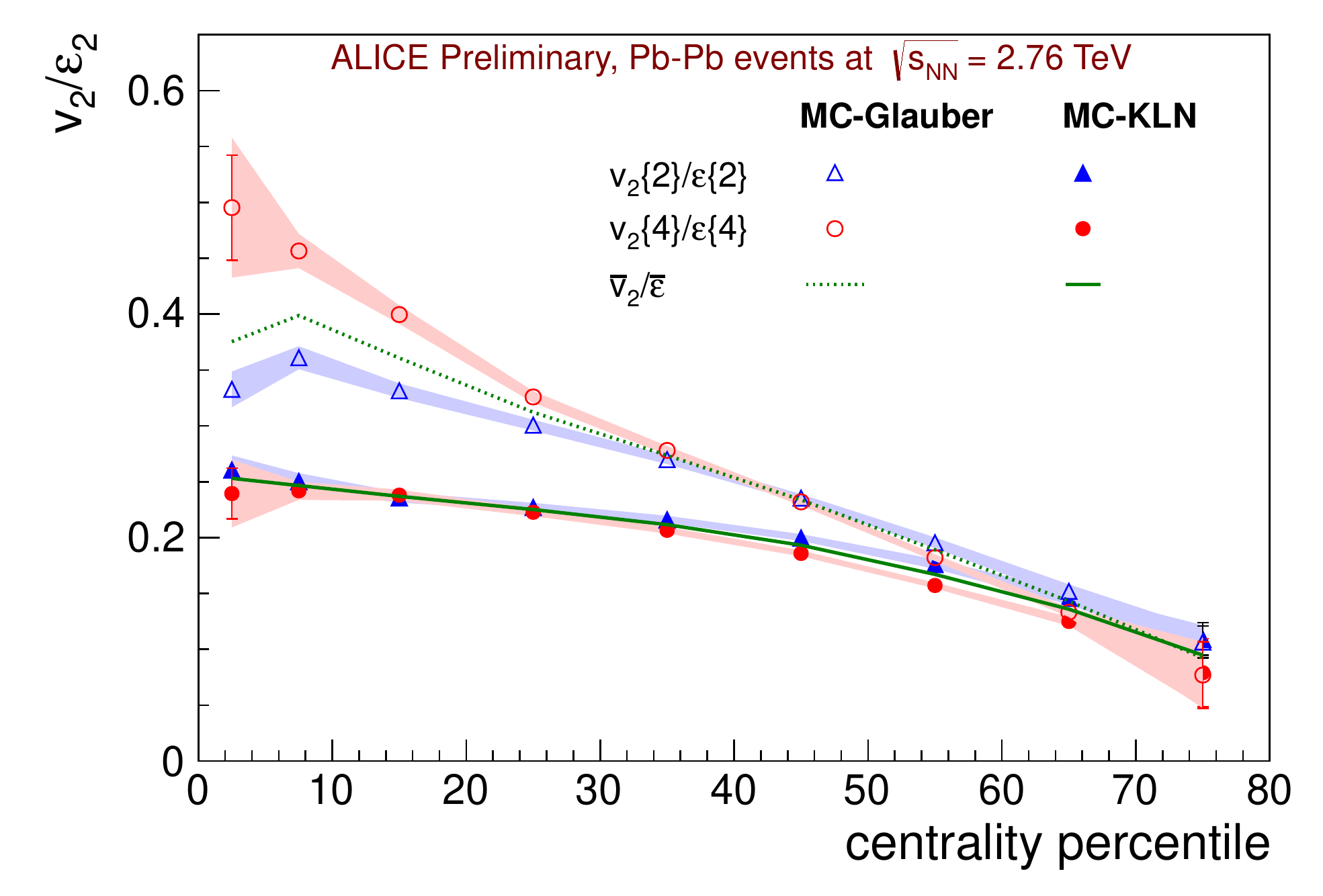}
\includegraphics[width=8cm]{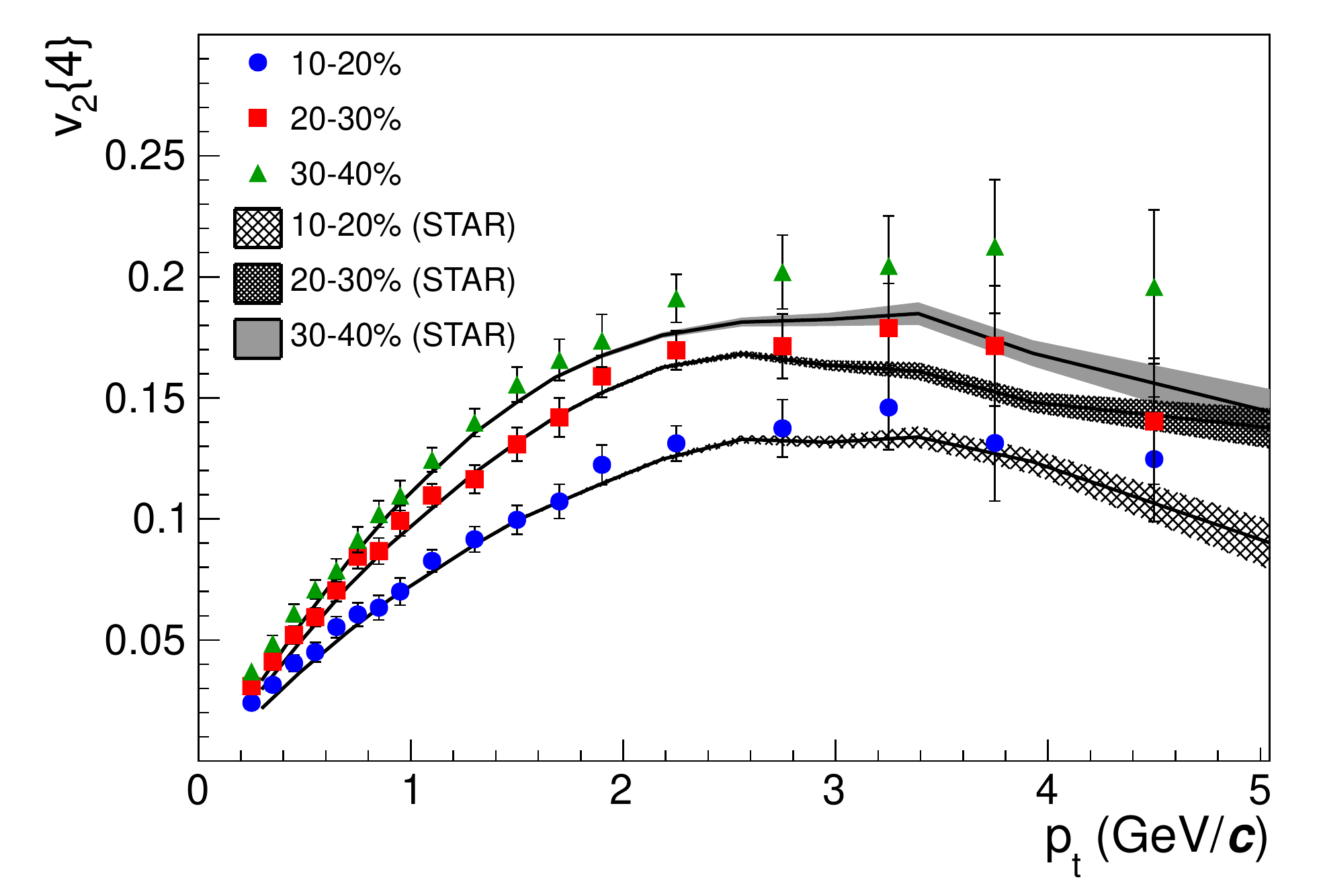}
\caption{
Left: The ratio of elliptic flow to eccentricity versus collision centrality.
Right: Elliptic flow for three centrality classes measured as function of transverse momentum $p_{\rm t}$. 
The bands indicate RHIC results measured by STAR. Figure taken from~\cite{Aamodt:2010pa}.}
\label{fig:figure4} 
\end{figure}
To eliminate the trivial dependence of $\varepsilon_2$ and $v_2$ on centrality and to increase the 
sensitivity to the properties of the medium it is illustrative to plot the ratio $v_2/\varepsilon_2$. 
In the left panel of Fig.~\ref{fig:figure4} we show the ratios $v_2\{2\}/\varepsilon_2\{2\}$ and $v_2\{4\}/\varepsilon_2\{4\}$ 
with $\varepsilon_2$ calculated from MC-KLN or MC-Glauber.
These two ratios should be identical, provided that the fluctuations are correctly described in the models, which is seen not to
be the case for the more central collisions in MC-Glauber.
To reduce the sensitivity to fluctuations we plot $\bar{v}_2/\bar{\varepsilon}_2$ as defined by Eq.~\ref{eq:flucts} (curves in Fig.~\ref{fig:figure4}). 
The mismatch between the two curves indicates a sensitivity to the initial conditions which still obscures the direct relation between $v_2/\varepsilon_2$ 
and the properties of the medium. 

\section{$p_{\rm t}$-differential Elliptic Flow}
Elliptic flow as a function of transverse momentum $p_{\rm t}$ is sensitive to the evolution and freeze-out conditions of the created system. 
The right panel of Fig.~\ref{fig:figure4} shows that the charged particle $p_{\rm t}$-differential elliptic flow, compared to RHIC,  
does not change within uncertainties at low $p_{\rm t}$~\cite{Aamodt:2010pa} which is remarkable because the beam energies 
differ by more than one order of magnitude. The 30\% increase in the integrated flow, shown in Fig.~\ref{fig:figure1}, 
must therefore be due to an increase in average transverse momentum.

\begin{figure}[t]
\includegraphics[width=8cm]{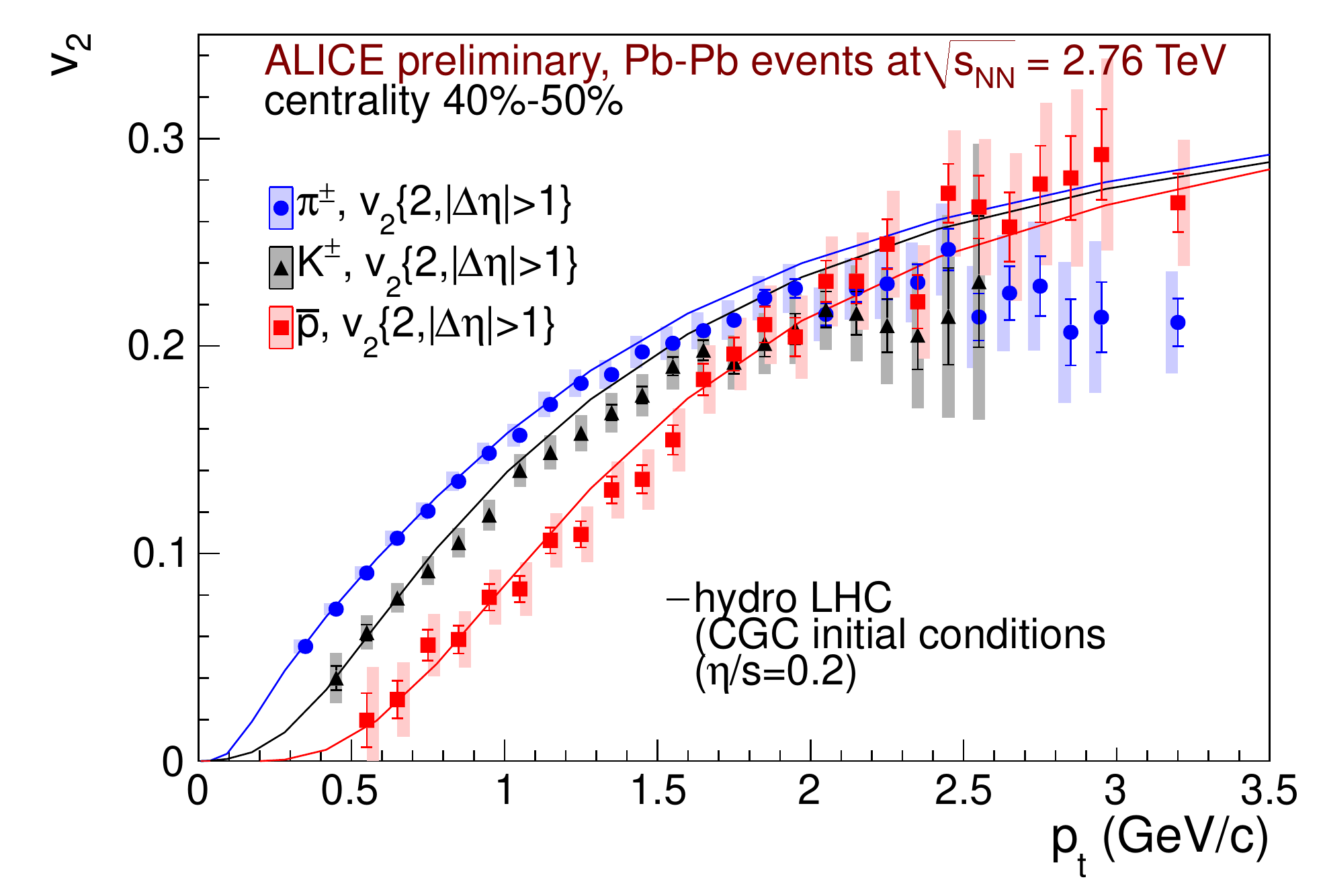}
\includegraphics[width=8cm]{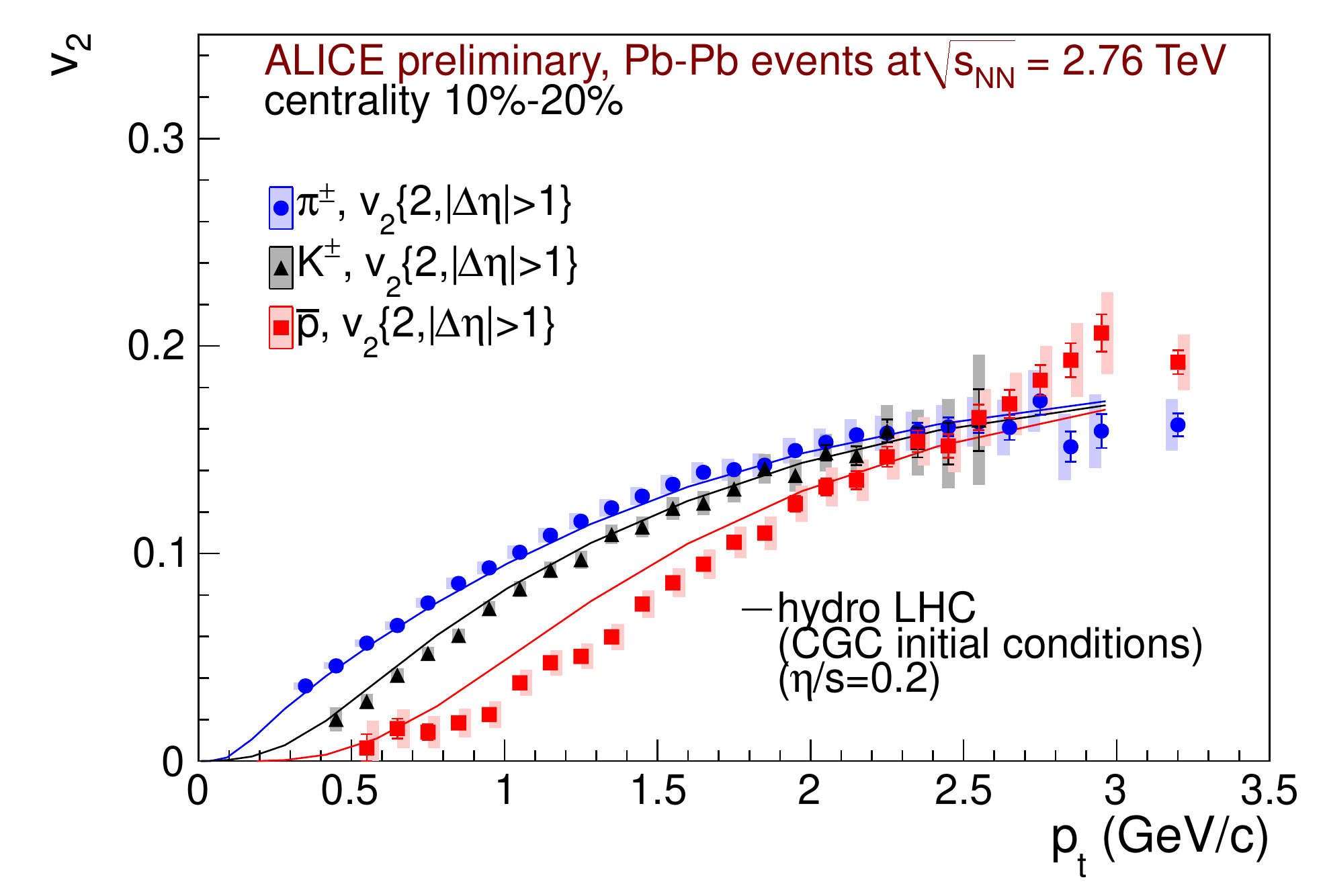}
\caption{
The $p_{\rm t}$-differential elliptic flow for pions, kaons and antiprotons for 40\%--50\% (left) and 
10\%--20\% (right) collision centrality. 
The curves are hydrodynamical model calculations. Figures taken from~\cite{ref:Mikolaj}.}
\label{fig:figure5} 
\end{figure}

In hydrodynamical model calculations this increase in mean $p_{\rm t}$ is due to a larger transverse flow at higher energies.
This leads to a more pronounced mass dependence of the elliptic flow.  
In Fig.~\ref{fig:figure5} we show identified particle $p_{\rm t}$-differential elliptic flow compared to hydrodynamic model predictions.
The hydrodynamic model predictions from~\cite{Shen:2011eg} (curves in Fig.~\ref{fig:figure5}) describe for 
mid-central collisions very well the measured 
$v_2$($p_{\rm t}$) for pions, kaons and antiprotons at  low $p_{\rm t}$ 
(left panel of Fig.~\ref{fig:figure5}).
For more central collisions (right panel) the hydrodynamical model predictions 
well describe the flow of pions and kaons but not that of the antiprotons.
This mismatch  is also observed for the spectra~\cite{ref:Michele} and may indicate a larger radial flow in the data.
At RHIC this was also observed and there a better description of the antiproton flow was obtained 
by introducing a hadronic cascade afterburner in the calculations~\cite{Teaney:2001av}.

\begin{figure}[h!]
\includegraphics[width=8cm]{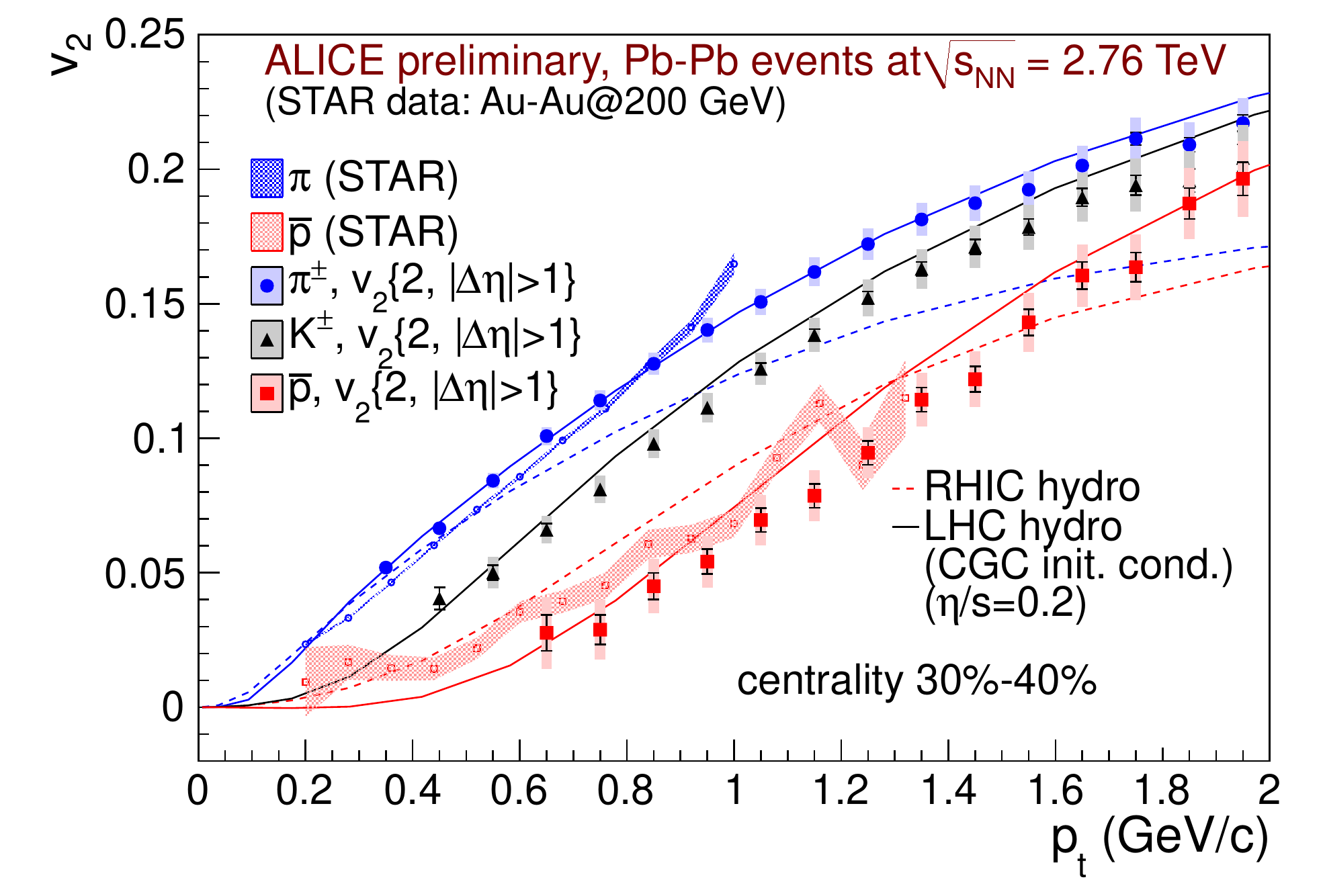}
\includegraphics[width=8cm]{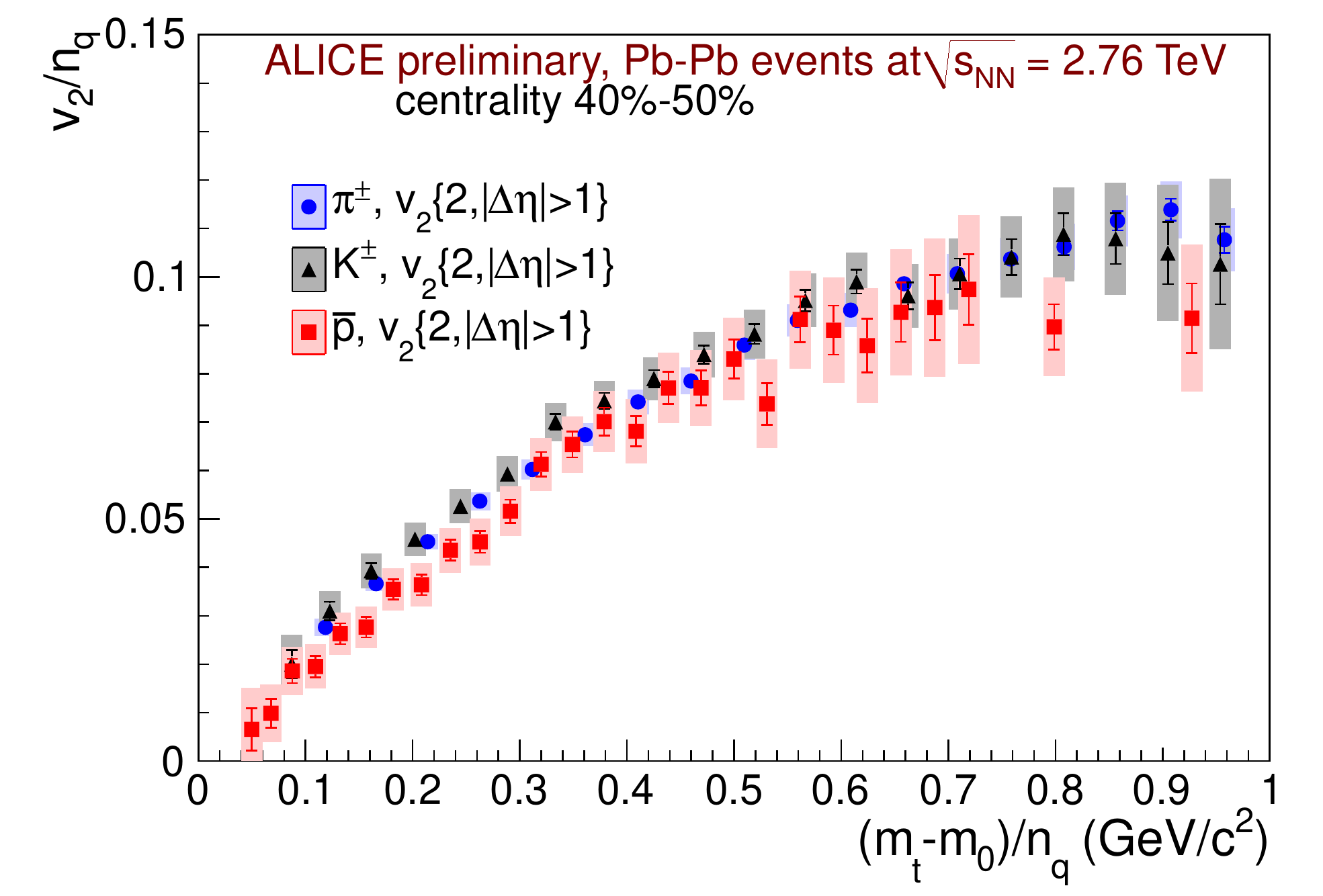}
\caption{
Left: The $p_{\rm t}$-differential elliptic flow for pions, kaons and antiprotons compared to results from RHIC (shaded bands)
and hydrodynamical calculations at RHIC (dashed curves) and LHC (full curves). 
Right: Elliptic flow versus transverse kinetic energy both divided by the number of constituent quarks.
Figures taken from~\cite{ref:Mikolaj}.
}
\label{fig:figure6} 
\end{figure}
The energy dependence of the mass splitting at low $p_{\rm t} < 2$ GeV/$c$ is shown in more detail in the left panel of Fig.~\ref{fig:figure6}.
To enable a direct comparison to the STAR pion and antiproton data (bands in the figure), 
the elliptic flow measured by ALICE is plotted for the centrality bin of 30\%--40\%.
We observe a small but significant increase with energy of the mass splitting between pions and antiprotons.
An increased mass splitting is also observed in hydrodynamical calculations between RHIC (dashed curves) and LHC energies (full curves) although
the antiproton elliptic flow is overestimated in both these calculations.

Hydrodynamics predicts that the mass splitting pattern persists at large values of $p_{\rm t}$ in contrast to what is observed in the 
data, as is apparent from Fig.~\ref{fig:figure5} above $p_{\rm t} \approx$~2~GeV/$c$. 
An elegant explanation of the particle type dependence and magnitude of $v_2$ at larger $p_{\rm t}$ 
is provided by the coalescence picture. At RHIC it was observed that $v_2 / n_{q}$ showed a universal scaling 
when plotted versus $(m_{\rm t} - m)/n_{q}$. Here $n_{q}$ is the number of 
constituent quarks and $m_{\rm t} = \sqrt{p_{\rm t}^2 + m^2}$ is the transverse mass, with 
$m$ the mass of the particle.
The right panel of Fig.~\ref{fig:figure6} shows this scaling at 2.76 TeV for 40\%-50\% collision centrality measured by ALICE.
The data show a clear scaling for pions and kaons but not for antiprotons. 

\begin{figure}[h!]
\includegraphics[width=8cm]{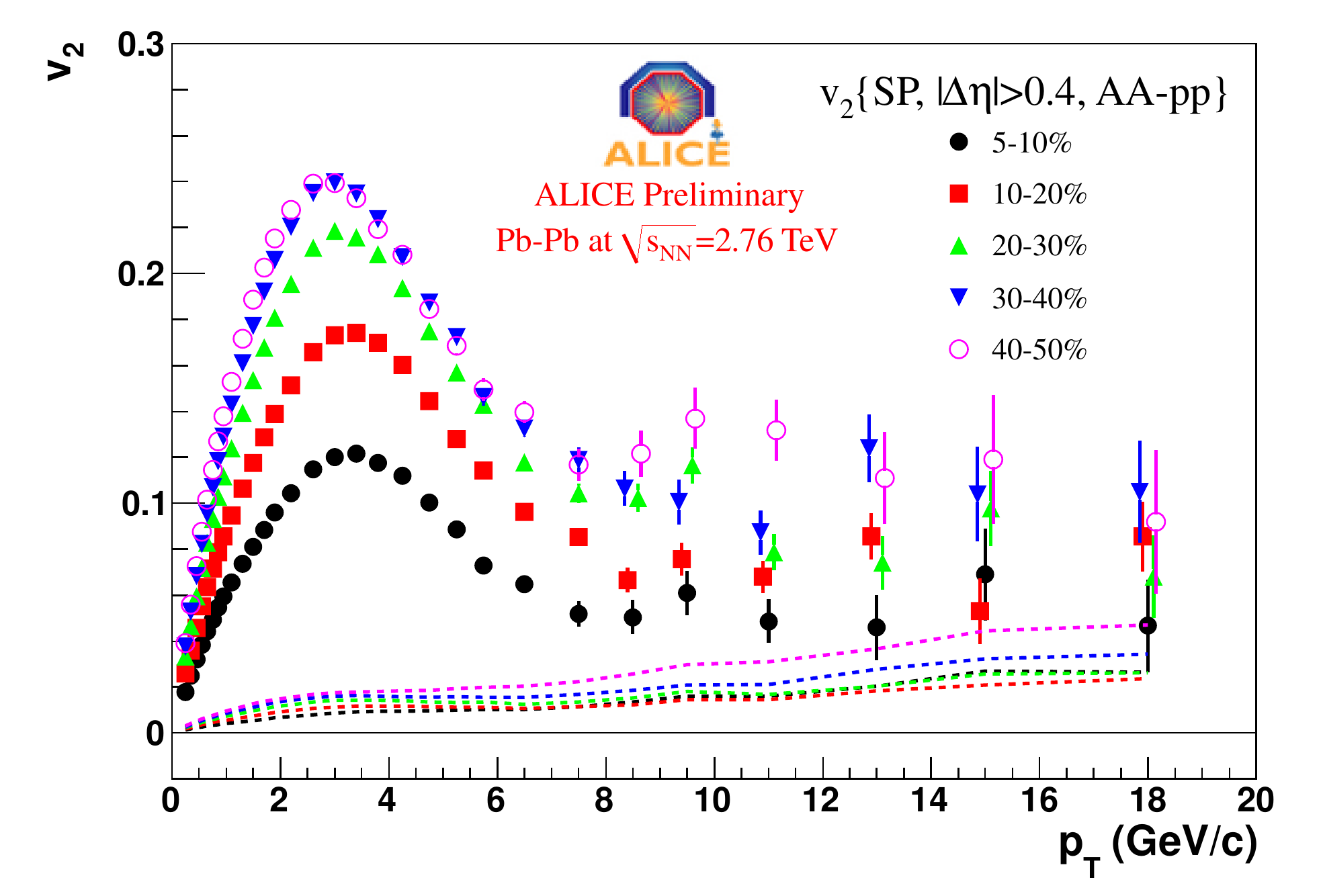}
\includegraphics[width=8cm]{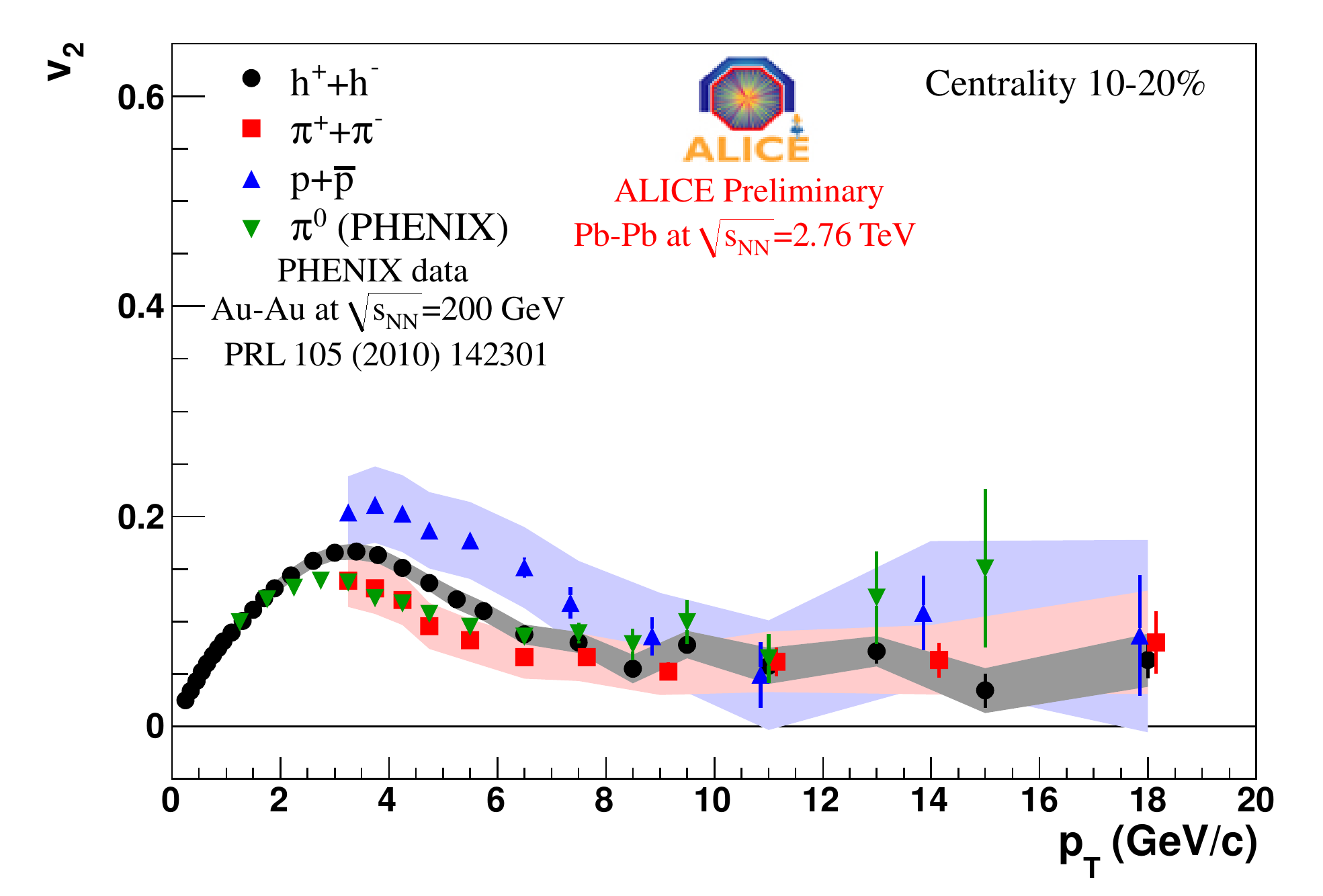}
\caption{
Left: Charged particle $p_{\rm t}$-differential elliptic flow for different collision centralities~\cite{ref:Alex}. 
Right: Identified particle elliptic flow at high-$p_{\rm t}$. For comparison the $\pi^0$ data measured by PHENIX are also shown.
Figures taken from~\cite{ref:Alex}.
}
\label{fig:figure7} 
\end{figure}
At sufficiently high transverse momentum, hadron yields are thought to contain a significant contribution  
from the fragmentation of high energy partons produced at initial hard scatterings. 
These high energy partons are predicted to lose energy when traversing nuclear matter. 
This energy loss is expected to depend strongly on the color charge density of the medium and on the path length traversed
by the propagating parton. 
Because this path length depends on the azimuthal emission angle with respect to the reaction plane, an azimuthal anisotropy of
particle emission is introduced at large $p_{\rm t}$. 
Indeed, significant values of $v_2$ are found between 8 and 20 GeV/$c$ as is shown in the left panel of Fig.~\ref{fig:figure7}.
This $v_2$ increases  from central to more peripheral collisions as is expected from the path length dependence of parton energy loss.
To investigate where the coalescence regime stops and where the parton energy loss mechanism might become dominant, we show in the
right panel of Fig.~\ref{fig:figure7} the identified particle $v_2$ at large $p_{\rm t}$. 
We see that up to about 8 GeV/$c$ the proton $v_2$ is larger than the pion $v_2$, as one would expect from coalescence. 

\section{Higher Harmonic Anisotropic Flow Coefficients}
Due to fluctuations in the initial matter distribution the initial spatial geometry has not a smooth almond shape but, instead,  a more 
complex  spatial geometry which may possess also odd harmonic symmetry planes. 
These are predicted to give rise to odd harmonics like triangular flow $v_3$.
Recently it was realized that these odd harmonics are particularly sensitive to both $\eta$/s and the initial conditions, which generated 
strong theoretical and experimental interest~\cite{:2011vk}.
\begin{figure}[h!]
\includegraphics[width=8cm]{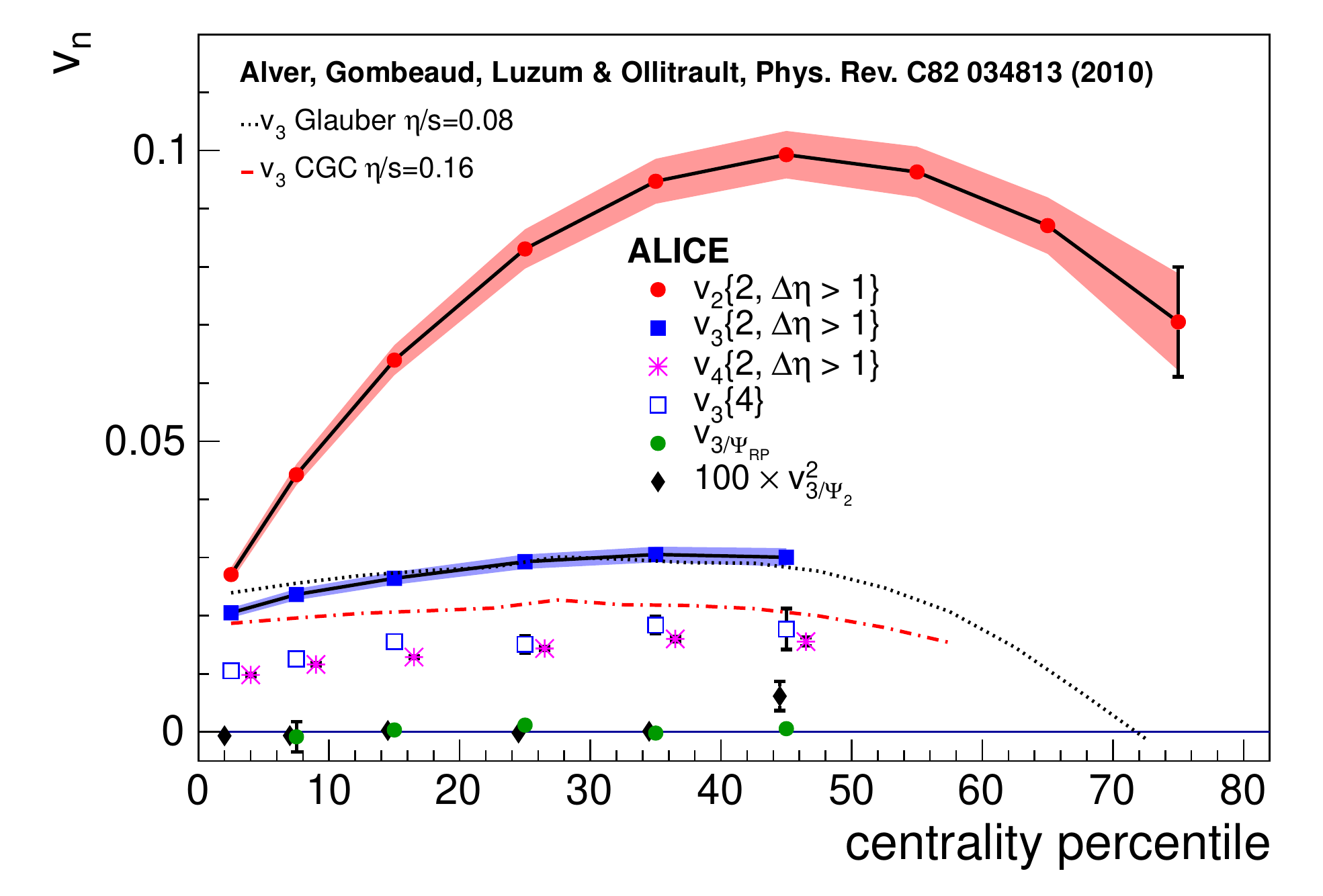}
\includegraphics[width=8cm]{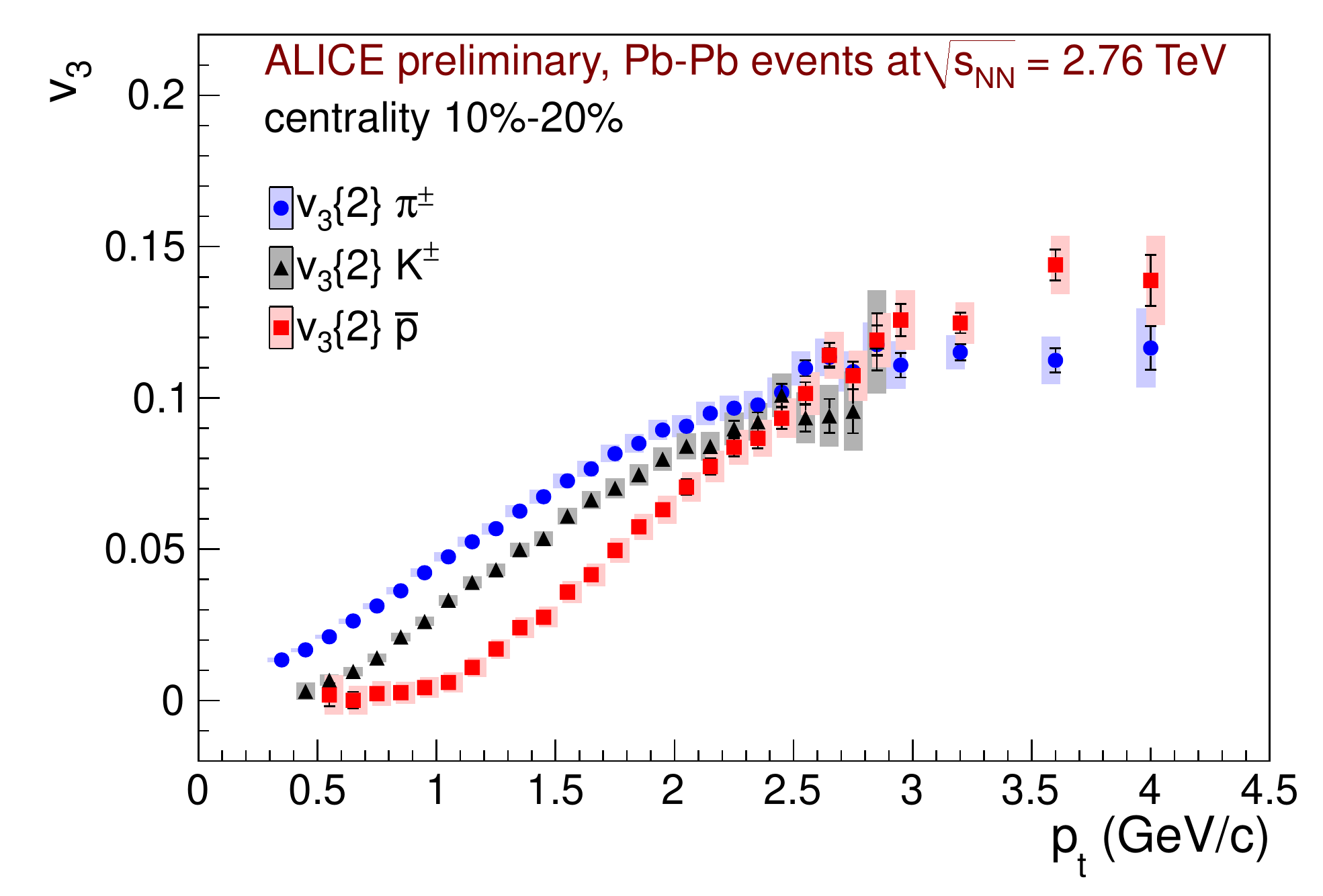}
\caption{
Left: Integrated $v_2$, $v_3$ and $v_4$, full and open squares show
$v_{3}\{2\}$ and $v_{3}\{4\}$ respectively. In addition we show $v_{3/\Psi_{2}}^{2}$ and $v_{3/\Psi_{\rm RP}}$, 
which represent the triangular flow measured relative to the second order event plane and the reaction plane, respectively.
The dashed curves are hydrodynamical predictions~\cite{Alver:2010dn} described in the text. 
Figure adapted from~\cite{:2011vk}
Right: The $p_{\rm t}$-differential triangular flow for pions, kaons and antiprotons. Figure taken from~\cite{ref:Mikolaj}.
}
\label{fig:figure8} 
\end{figure}
The left panel of Fig.~\ref{fig:figure8} shows that $v_3$ is indeed significant 
and does not depend strongly on centrality. The magnitude and centrality dependence of $v_3$ is 
reasonably well described by 
predictions from a hydrodynamic model calculation with Glauber initial conditions
and $\eta/s = 0.08$ (dotted curve), in contrast to a calculation based on MC-KLN CGC initial conditions with $\eta/s = 0.16$,  
which under-predicts the data (dashed dotted curve). 
This suggests that the value of $\eta/s$ for the matter created in these collisions is small. 
However, currently there is no model calculation which
can describe both the integrated and $p_{\rm t}$-differential $v_2$ and $v_3$ at LHC energies with one set of parameters.
To investigate further its hydrodynamic origin, the $p_{\rm t}$-differential $v_3$ has been 
measured for pions, kaons and antiproton, as is shown in the right panel of Fig.~\ref{fig:figure8}. 
It is seen that the mass splitting pattern in elliptic flow is clearly present in triangular flow as well. 
In addition to providing constraints on $\eta/s$ and the initial conditions, 
it is shown in~\cite{:2011vk} that the measured $v_2$ and $v_3$ flow coefficients also provide a natural 
explanation for the two-particle correlations structures--- the so-called Mach cone and soft ridge--- first seen at RHIC and 
later at the LHC~\cite{ref:janfiete}.  

\section{Summary}

In this overview we have presented first results on anisotropic flow measured in Pb--Pb collisions 
at $\sqrt{s_{_{\rm NN}}} = 2.76$ TeV by ALICE at the LHC. 
Details on these measurements can be found elsewhere in these proceedings~\cite{ref:Ante,ref:Mikolaj,ref:Alex,ref:Ilya}.
We find that for transverse momenta of up to about 2 GeV/$c$ hydrodynamical model calculations 
can give a good description of the measurements, even though currently there is no simultaneous description 
of all the anisotropic flow data with a single choice of initial conditions and $\eta/s$.
The stronger elliptic flow, compared to RHIC, shows that the system created at LHC collision energies still behaves like an almost 
perfect fluid. The first measurements of the higher harmonic anisotropic flow coefficients, in particular $v_3$, already provide new strong 
constraints on $\eta/s$ and the initial conditions.   
Barely six months after the first lead on lead collisions, the ALICE anisotropic flow measurements show that the properties of the created matter 
at the LHC can be studied with unprecedented precision.


\section*{References}
\begin{thebibliography}{99}

\bibitem{Voloshin:2008dg}
  S.~A.~Voloshin, A.~M.~Poskanzer and R.~Snellings,
  in Landolt-Boernstein, {\em Relativistic Heavy Ion Physics}, Vol. 1/23, p 5-54 (Springer-Verlag, 2010)
 
\bibitem{Aamodt:2010pa}
  K.~Aamodt {\it et al.}  [ALICE Collaboration],
  Phys.\ Rev.\ Lett.\  {\bf 105}, 252302 (2010)
 
\bibitem{Voloshin:1994mz}
  S.~Voloshin and Y.~Zhang,
  Z.\ Phys.\  C {\bf 70} 665 (1996)

\bibitem{Ollitrault:1992bk}
 J.~Y.~Ollitrault,
 Phys.\ Rev.\  D {\bf 46}, 229 (1992)
 
 \bibitem{Kovtun:2004de}
  P.~Kovtun, D.~T.~Son, A.~O.~Starinets,
  Phys.\ Rev.\ Lett.\  {\bf 94}, 111601 (2005)

\bibitem{ref:Ante}{A.~Bilandzic, These proceedings.}

\bibitem{Alver:2007qw}
  B.~Alver {\it et al.} [PHOBOS Collaboration],
  Phys.\ Rev.\ Lett.\  {\bf 104}, 142301 (2010)
  
\bibitem{Drescher:2007ax}
  H.~-J.~Drescher, Y.~Nara,
  Phys.\ Rev.\  {\bf C76}, 041903 (2007)

\bibitem{Miller:2007ri}
  M.~L.~Miller, K.~Reygers, S.~J.~Sanders, P.~Steinberg,
  Ann.\ Rev.\ Nucl.\ Part.\ Sci.\  {\bf 57}, 205-243 (2007)

\bibitem{ref:Mikolaj}{M.~Krzewicki, These proceedings.} 

\bibitem{Shen:2011eg}
  C.~Shen, U.~W.~Heinz, P.~Huovinen, H.~Song,
  [arXiv:1105.3226 [nucl-th]].

\bibitem{ref:Michele}{M.~Floris, These proceedings.} 

\bibitem{Teaney:2001av}
  D.~Teaney, J.~Lauret, E.~V.~Shuryak,
  [nucl-th/0110037]

\bibitem{ref:Alex}{A.~Florin~Dobrin, These proceedings.}

\bibitem{:2011vk}
  K.~Aamodt {\it et al.}  [ALICE Collaboration],
  arXiv:1105.3865 [nucl-ex].
 
\bibitem{Alver:2010dn}
  B.~H.~Alver, C.~Gombeaud, M.~Luzum, J.~-Y.~Ollitrault,
  Phys.\ Rev.\  {\bf C82}, 034913 (2010)
  
\bibitem{ref:janfiete}{J. F. Grosse-Oetringhaus, These proceedings.}

\bibitem{ref:Ilya}{I.~Selyuzhenkov, These proceedings.}

\end{thebibliography}
\end{document}